# The equations of medieval cosmology

*Roberto Buonanno and Claudia Quercellini*
*University of Rome "Tor Vergata"*
*Department of Physics*
*Via della Ricerca Scientifica 1, 00133 Rome, Italy*

**Abstract.** In Dantean cosmography the Universe is described as a series of concentric spheres with all the known planets embedded in their rotation motion, the Earth located at the centre and Lucifer at the centre of the Earth. Beyond these "celestial spheres", Dante represents the "angelic choirs" as other nine spheres surrounding God. The rotation velocity increases with decreasing distance from God, that is with increasing Power (Virtù). We show that, adding Power as an additional fourth dimension to space, the modern equations governing the expansion of a closed Universe (i. e. with the density parameter $\Omega_0 > 1$) in the space-time, can be applied to the medieval Universe as imaged by Dante in his Divine Comedy. In this representation the Cosmos acquires a unique description and Lucifer is not located at the centre of the hyperspheres.

The Dantean cosmography started about 600 years ago, when the medieval concept of the Universe was already fading away.

It is probably not a mere coincidence that the first scholars to deal with the problem at the end of XVth century were two architects who accepted the challenge to visualise those sites, their shapes and sizes which had been poetically described by Dante in his Divine Comedy.

The first one was a young Filippo Brunelleschi, who analysed the "*site and the size*" (*il sito e le misure*) of all the locations reported in the Divine Comedy[i]; the second one, Antonio di Tuccio Manetti[ii,iii] a specialist in perspective studies, derived the shape and size of the Hell examining Dante's poem.

Even if both the architects never published their studies, their conclusions are at the basis of Hell's iconography which can found in the XVIth century editions of the Divine Comedy[iv].

This iconography, which continues nowadays, represents the Hell as a cone formed when Lucifer fell from the Sky. The apex of the cone is at the centre of the Earth, where Lucifer remained jammed, with his head and feet respectively in the boreal and austral hemisphere.

The issue was so important in the XVIth century that even a young Galileo Galilei got involved and in 1587 he delivered two lectures on the Dante's Hell at the Accademia Fiorentina[v]. By a close scrutiny of Dante's verses, in these lectures Galileo worked out the size of the Hell, as well as Lucifer's.

Although the Hell has been abundantly illustrated for six centuries, it turns out that this is not the case for the Paradise. To our knowledge, Michelangelo Cactani[vi] in 1855 was the first to publish a series of woodcuts specifically intended to illustrate Dante's cosmography, Paradise included. It is noticeable that all the ensuing representations faithfully trace Cactani's scheme (figure 1).

In the Divine Comedy Dante figures the Empyrean in a specular manner with respect to the celestial spheres. In the Empyrean the celestial spheres correspond to nine "angelic choirs" orbiting around God with velocities that increase with decreasing distance from the center[vii].

At first sight, this is the opposite of what happens for the celestial spheres whose velocities increase with increasing the distance from their center of rotation, the Earth. However, Dante clarifies that the parameter which drives the rotation speed is, in both cases, the Power, i.e. the distance from God, called by Dante "Virtù"[viii].

Therefore all the spheres of the Universe, both the celestial and those of the Empyrean, rotate with velocities that increase with decreasing the distance from God.

It has been already recognized that, in order to represent the whole medieval Universe (Earth, planets, Hell, Empyrean and their rotation), Dante was forced to add a fourth spatial dimension to the usual three[ix,x].

In this work we have investigated whether the equations of the space-time could be translated into the Dantean space-power.

*Dynamics of the expansion of the Universe (Newtonian cosmology)*
For the time being it is not necessary to make use of General Relativity. Let us consider the expansion of the Universe as it can be derived in Newtonian cosmology:

$$\ddot{a}(t) = -\frac{4\pi G}{3}\rho(t)a(t)$$

or equivalently

$$\dot{a}^2 = \frac{8\pi G}{3}\rho(t)a^2(t) - Kc^2$$

where $a(t)$ is the cosmic scale factor, $\rho(t)$ is the mass density, $G$ is the gravitational constant and $Kc^2$ is an integration constant.

With a coordinate transformation from space-time to space-power and substituting $a$ with $R$. It follows that

$$\frac{\Delta R}{\Delta V} = \frac{1}{R^{1/2}}C_1 - C_2$$

where $R(V)$ is the radius of the celestial spheres and the angelic choirs as a function of the power $V$.

The Dantean Universe in the space-power manifestly corresponds to a closed Universe in the space-time ($K>0$ case), for which the parametric solutions are the following

$$\begin{cases} R(\theta) = A(1-\cos\theta) + R_L \\ V(\theta) = B(\theta - \sin\theta) \end{cases}$$

where $A$ and $B$ are constants and $R_L$ is minimum size of the Universe in the space-power.

The latter, according to Dante's view, is the size of Lucifer, which is in fact located by the poet at the center of the Earth. Galileo Galilei has shown that it is possible to derive this size from Dante's verses obtaining $R_L = 645$ braccia[xi].

The constant A can be related to the maximum radius of the Universe, which is the radius of the Primum Mobile, $R_{PM}$, for which we assumed that the power holds $V=1$.

The equations for the Universe as conceived by Dante are then

$$\begin{cases} R(\theta) = \frac{R_{PM}-R_L}{2}(1-\cos\theta) + R_L \approx \frac{R_{PM}}{2}(1-\cos\theta) + R_L \\ V(\theta) = \frac{K(\theta)}{\pi}(\theta - \sin\theta) \end{cases}$$

with the boundary conditions $R(\pi)=R_{PM}$, $V(\pi)=1$; $R(0)=R_L$, $V(0)=0$ and $0 \leq \theta \leq 2\pi$.

Here $K(\theta)$ is an arbitrary function which allows to increase the power of the angelic choirs and obeys to the condition $K(\pi)=1$.

This system, which is tuned on the power of the celestial spheres of the planets and their distance from the Earth, predicts the power of the angelic choirs once that their distance from God is known.

It is certainly an astonishing outcome that a medieval cosmological model can be naturally represented by a set of equations worked out centuries apart.

An insight into this evidence is offered by the two following arguments which are often ignored by historians: first, the Dantean system is completely original and can be barely superimposed to the ptolemaic scheme and, second, Dante must embody a Universe which is not in contradiction neither with his astronomical knowledge nor with his religious view.

These arguments forced him to realize that the traditional space, i.e. that of the celestial spheres, is embedded in a quadri-dimensional space, where the fourth dimension is the power. As a consequence, God results to be in the centre of the space-power, just as the Big Bang is the centre of the space-time, even though, in the three-dimensional space, they appear to be at the edge of the Universe.

---

[i] G. Vasari, *Le Vite de' più eccellenti pittori scultori e architetri,* Siena, 1791

[ii] C. Landino, *Comento sopra la Comedia,* Firenze, Nicolò di Lorenzo della Magna, 1481

[iii] G. Benivieni, *Commedia di Dante insieme con un dialogo circa el sito, forma et misure dello Inferno,* Filippo di Giunta, Firenze, 1506

[iv] *Ibidem*

[v] G. Galilei, *Due Lezioni all'Accademia Fiorentina, in Scritti letterari,* Firenze, Le Monnier, 1970

[vi] M. Cactani, *La materia della Divina Commedia di Dante Alighieri dichiarata in VI tavole,* 1855, folio.

[vii] D. Alighieri, *La Divina Commedia, Paradiso* XXVIII, 25

[viii] D. Alighieri, *La Divina Commedia, Paradiso,* XXVIII, 64

[ix] R. Osserman, *A Mathematical Exploration of the Cosmos*, Anchor Books, 1995

[x] H. R. Patapievici, *Gli occhi di Beatrice,* Bucarest, Humanitas Editrice, 2004

[xi] G. Galilei, *Due lezioni all'Accademia Fiorentina, in Scritti letterari,* Firenze, Le Monnier, 1970

[xii] D. Alighieri, *Convivio,* Trattato II, cap. III, http://www.filosofico.net/conviviodante.htm

*Fig. 1: The Dantean Universe according to M. Cactani, 1855.*
*Following the ptolemaic cosmology, the Earth is immovable at the centre of the Universe and surrounded by the celestial spheres which carry the known planets.*
*Beyond the planet sphere there is the sphere of fixed stars and, beyond it, the Primum Mobile, which transmits the rotation motion to all the spheres[xii]. Since the Earth does not have rotation motion, all the celestial spheres must have a 24 hours revolution motion around the Earth with the same angular velocity (except the Moon sphere which is slower by about 13degree per day). It follows that the linear velocity of spheres increases with the distance from the Earth.*

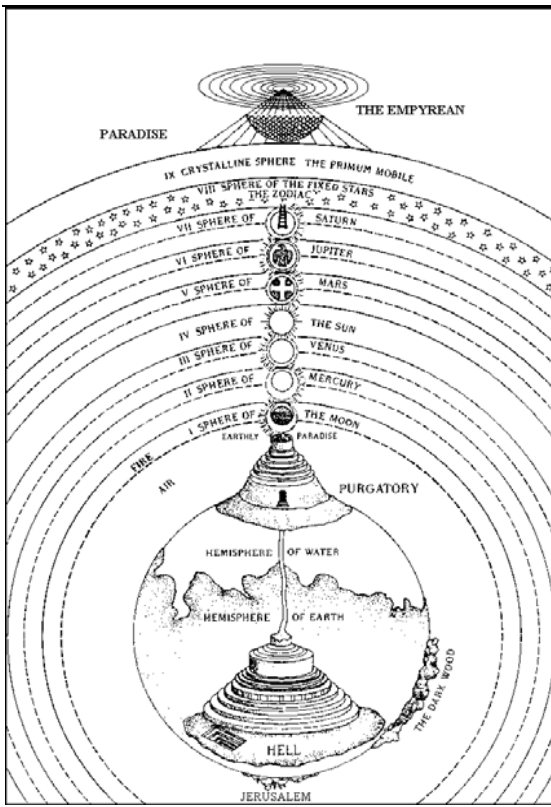